\documentclass[aps,prd,twocolumn,showpacs,preprintnumbers,nofootinbib,amsmath,amssymb]{revtex4}

\usepackage{graphicx}
\usepackage{dcolumn}
\usepackage{bm}
\usepackage{amsmath}
\usepackage{braket}
\usepackage[normalem]{ulem}
\usepackage[hidelinks]{hyperref}
\usepackage{epstopdf}
\usepackage{placeins}
\usepackage{multirow}
\usepackage{makecell}
\usepackage[normalem]{ulem} 

\newcommand{\beq}{\begin{eqnarray}}
\newcommand{\eeq}{\end{eqnarray}}
\newcommand{\beqnn}{\begin{eqnarray*}}
\newcommand{\eeqnn}{\end{eqnarray*}}
\newcommand{\Tr}{\ensuremath{\mathrm{Tr}}}
\newcommand{\round}{\ensuremath{\mathrm{round}}}
\newcommand{\YM}{\mathit{YM}}

\def\spose#1{\hbox to 0pt{#1\hss}}
\def\ltapprox{\mathrel{\spose{\lower 3pt\hbox{$\mathchar"218$}}
 \raise 2.0pt\hbox{$\mathchar"13C$}}}

\begin{document}

\title{Large-$N$ $SU(N)$ Yang--Mills theories
with milder topological freezing}

\author{Claudio Bonanno}
\email{claudio.bonanno@pi.infn.it}
\affiliation{Universit\`a di Pisa and INFN Sezione di Pisa,\\ 
	Largo Pontecorvo 3, I-56127 Pisa, Italy,\\
	Centro Nazionale INFN di Studi Avanzati GGI,\\
	Largo Enrico Fermi 2, I-50125 Firenze, Italy}

\author{Claudio Bonati}
\email{claudio.bonati@unipi.it}
\affiliation{Universit\`a di Pisa and INFN Sezione di Pisa,\\ 
Largo Pontecorvo 3, I-56127 Pisa, Italy
}

\author{Massimo D'Elia}
\email{massimo.delia@unipi.it}
\affiliation{Universit\`a di Pisa and INFN Sezione di Pisa,\\ 
Largo Pontecorvo 3, I-56127 Pisa, Italy
}

\date{\today}

\begin{abstract}
We simulate $4d$ $SU(N)$ pure-gauge theories at large $N$ using a parallel
tempering scheme that combines simulations with open and periodic boundary
conditions, implementing the algorithm originally proposed by 
Martin Hasenbusch for $2d$
$CP^{N-1}$ models. That allows to dramatically suppress the
topological freezing suffered from standard local algorithms, reducing the
autocorrelation time of $Q^2$ up to two orders of magnitude. Using this
algorithm in combination with simulations at non-zero imaginary $\theta$ we are
able to refine state-of-the-art results for the large-$N$ behavior of the
quartic coefficient of the $\theta$-dependence of the vacuum energy $b_2$,
reaching an accuracy comparable with that of the large-$N$ limit of the
topological susceptibility.
\end{abstract}

\maketitle

\section{Introduction}
\label{sec:intro}

The dependence of physical observables on the topological parameter
$\theta$ is one of the most interesting properties of four dimensional $SU(N)$
pure-gauge theories. The parameter is 
coupled in the action to the topological charge
\beq
\label{eq:cont_def_topocharge}
Q = \hspace{-2pt} \int d^4 x \, q(x)
=\frac{1}{64\pi^2} \varepsilon_{\mu\nu\rho\sigma}\hspace{-2pt} \int d^4x\, F^a_{\mu\nu}(x) F^a_{\rho\sigma}(x) \, ;
\eeq
$\theta$-dependence 
can be studied to achieve a better understanding of the
non-perturbative features of Yang--Mills
theories, but also has direct
phenomenological implications for hadron physics in the limit of large number of
colors~\cite{tHooft:1973alw,tHooft:1976rip,Witten:1978bc,Witten:1979vv,Veneziano:1979ec,Witten:1980sp,DiVecchia:1980yfw}.

A particularly interesting quantity, whose dependence on $\theta$ has been
thoroughly investigated, is the vacuum energy density $E(\theta)$, which is 
formally defined by the relation
\beq\label{eq:free_energy_def}
E(\theta) \equiv -\frac{1}{V} \log \int [dA] e^{-S_{\YM}[A]+i\theta Q[A]}\ ,
\eeq
where $V$ is the four-dimensional euclidean space-time volume. The functional
form of $E(\theta)$ is known only for very specific theories, like for QCD close
to the chiral limit~\cite{DiVecchia:1980yfw}. It is customary to consider a Taylor
expansion of $E(\theta)$ around $\theta=0$. Since $Q$ is odd under a
$\mathrm{CP}$ transformation, only even powers of $\theta$ appear in the
expansion, which can be parametrized in the form
\beq\label{eq:vacuum_energy_theta_dep_parametrization}
E(\theta) -E(0) = \frac{1}{2} \chi \theta^2 \left( 1 + \sum_{n=1}^{\infty} b_{2n} \theta^{2n} \right)\ .
\eeq
The coefficients of this expansion are related to the cumulants of the
topological charge distribution at $\theta=0$, for instance
$\chi=\langle Q^2\rangle_{\theta=0}/V$, where $\chi$ is the 
topological susceptibility.

While the exact numerical values of the coefficients $\chi$ and $b_{2n}$ are
generically unknown, something is known about their dependence on the number
of colors $N$, at least when $N$ is large enough. Indeed, assuming that a non-trivial $\theta$-dependence is present in the large-$N$ limit, it is 
possible to fix the large $N$ scaling form of the energy 
density $E(\theta,N)=N^2 \bar{E}(\theta/N)$~\cite{Witten:1979vv,Witten:1980sp,Witten:1998uka}, 
implying
\beq\label{eq:large_N_scaling}
\chi   &=& \bar{\chi} + O\left(\frac{1}{N^2}\right),\\
b_{2n} &=& \frac{\bar{b}_{2n}}{N^{2n}} + O\left(\frac{1}{N^{2\left(n+1\right)}}\right)\ .
\eeq
The value of $\bar{\chi}$ is related to the mass of the $\eta^\prime$ meson by the
Witten--Veneziano formula~\cite{Witten:1979vv,Veneziano:1979ec}, which provides
the estimate $\bar{\chi} \simeq (180\,\text{MeV})^4$. Analytic estimates of
$\bar{\chi}$ and of the  $\bar{b}_{2n}$ coefficients are available only for two
dimensional models~\cite{DAdda:1978vbw, Campostrini:1991kv, DelDebbio:2006yuf,
Rossi:2016uce, Bonati:2016tvi}. 
Given the non-perturbative nature of $\theta$-dependence, the numerical lattice
approach is the natural tool to investigate such topics
quantitatively, and in particular to test large-$N$
predictions~\cite{Alles:1996nm, Alles:1997qe, DelDebbio:2004ns, DelDebbio:2002xa, DElia:2003zne,
DelDebbio:2006yuf, Lucini:2004yh,
Giusti:2007tu, Vicari:2008jw, Panagopoulos:2011rb, Ce:2015qha, Ce:2016awn,
Bonati:2015sqt, Bonati:2016tvi, Bonati:2018rfg, Bonati:2019kmf}. There are
however some non-trivial computational challenges that
have to be faced, especially in the large $N$
regime.

The first problem is related to the measure of the coefficients $b_{2n}$.
The task is challenging by itself since, contrary to what happens for $\chi$,
these coefficients approach zero as $N$ is increased. In addition, 
the simplest estimators available for $\theta = 0$ simulations
are based on the cumulants of the topological charge 
distribution,
which are not
self-averaging quantities, leading to a bad signal-to-noise ratio
for large volumes. Such problem can be solved by introducing 
an explicit source term in the action, coupled 
to the topological charge density, which corresponds in practice to 
an imaginary $\theta$ term~\cite{Bhanot:1984rx,Azcoiti:2002vk,Alles:2007br,Imachi:2006qq,Aoki:2008gv,Panagopoulos:2011rb,
Alles:2014tta,DElia:2012pvq,DElia:2013uaf,DElia:2012ifm}.
The method was exploited for the determination of the $b_{2n}$ 
coefficients first in Ref.~\cite{Panagopoulos:2011rb} and later developed and applied in several works
to both $SU(N)$ Yang--Mills theory \cite{Bonati:2015sqt, Bonati:2016tvi,
Bonati:2018rfg, Bonati:2019kmf} and two dimensional $CP^{N-1}$
models~\cite{Bonanno:2018xtd,Berni:2019bch}. 

The second problem is the so-called ``freezing problem'', 
it represents a well known 
problem in a wide range of theories sharing the presence of 
topological excitations~\cite{csd_full_QCD_delia, 
deForcrand:1997yw,Lucini:2001ej,DelDebbio:2002xa,Leinweber:2003sj,
DelDebbio:2004xh,
Luscher:2011kk,
Laio:2015era, Flynn:2015uma,
theta_dep_QCD_N_f_2+1,
Bonati:2017woi,Athenodorou:2020ani}
and it will be the
main topic of this work. When adopting local update algorithms on lattices with
periodic boundary conditions, the topological modes experience a severe
critical slowing down when approaching the continuum limit, with the
autocorrelation time of the topological charge which grows approximately
exponentially as a function of $1/a$~\cite{DelDebbio:2002xa,DelDebbio:2004xh}.
As a consequence gauge configurations stay fixed (frozen) in a given
topological sector for an exceedingly large amount of Monte Carlo time, thus
preventing a correct sampling of the path-integral distribution. This problem
becomes even worse in the large-$N$ limit, since at fixed lattice spacing $a$ the
autocorrelation time of the topological charge seems to grow exponentially with
the number of colors~\cite{DelDebbio:2002xa,DelDebbio:2004xh,Bonanno:2018xtd}.

Despite the fact that a definitive solution to the topological freezing problem
has been obtained only in toy models~\cite{Bonati:2017woi}, several strategies
have been proposed to reduce its severeness, in particular by reducing the
exponential critical slowing down to a polynomial critical slowing
down~\cite{Vicari:1992jy, Luscher:2011kk, Laio:2015era, Bonati:2017woi}, or to
extract information even from completely frozen configurations~\cite{Bietenholz:2015rsa}.  A popular method, proposed in
Ref.~\cite{Luscher:2011kk}, is to adopt open boundary conditions for the gauge
fields in the time direction instead of the usual periodic ones. The presence
of an open boundary eliminates barriers among different topological sectors
even in the continuum, and one expects the critical slowing down of the
topological modes to be essentially diffusive, with an autocorrelation time
increasing as $1/a^2$ approaching the continuum limit.  Using this method,
however, one also breaks translation invariance and loses completely any notion
of global topological charge. Therefore $\chi$ and $b_{2n}$ can only be
estimated from the integral of the $(2n+2)$-point connected correlators of the
topological charge density on the bulk of the lattice.

A different strategy, that keeps the advantages of the open boundary conditions
without breaking translation invariance, has been proposed by M.~Hasenbusch in
Ref.~\cite{Hasenbusch:2017unr}, where it was tested for two dimensional
$CP^{N-1}$ models. The basic idea of this method is to combine periodic and
open boundary conditions in a parallel tempering framework, using the copies
with open or partially open boundary conditions as sources of topological
fluctuations for the copy with periodic boundary conditions, which is the one on which
measures are performed. To reduce the number of copies to be used in the
parallel tempering, open boundary conditions are not enforced along all the
temporal boundaries, but only in a limited space region, that will be referred
to as the ``defect'' in the following. 

In Ref.~\cite{Berni:2019bch} it was shown that in two dimensional $CP^{N-1}$
models, for large $N$ values, the adoption of the Hasenbusch algorithm in
combination with the imaginary-$\theta$ method allows to achieve impressive
improvements compared to previously available results for the
$\theta$-dependence of these models. The aim of the present work is to test the
same setup in the case of four dimensional $SU(N)$ Yang--Mills theories at zero
temperature, comparing its performance with that of the standard simulations.
In doing this we will also refine the state-of-the-art results about the large-$N$
behavior of the $b_2$ coefficient.

This paper is organized as follows: in Sec.~\ref{sec:lattice_setup} we discuss
our lattice setup, along with the parallel tempering algorithm and the
imaginary-$\theta$ method, in Sec.~\ref{sec:results} we show the numerical
results obtained with the parallel tempering and finally in
Sec.~\ref{sec:conclusion} we draw our conclusions.

\section{Lattice setup}
\label{sec:lattice_setup}

In this section we introduce the discretizations adopted for the action and the
topological charge, we present a summary of the imaginary-$\theta$ method and
discuss the parallel tempering algorithm employed in our simulations.

\subsection{Lattice action and lattice topological charge}

We discretize the Yang--Mills action on an hyper-cubic lattice of size $L$
and with periodic boundary conditions in every direction (see Sec.~\ref{sec:defect}
for the defect) using the standard Wilson action:
\beq\label{eq:lat_def_action}
S_W = -\frac{\beta}{N} \sum_{x,\mu>\nu} \Re \Tr \left\{ \Pi_{\mu\nu}(x) \right\},
\eeq
where $\Pi_{\mu\nu}(x) \equiv
U_\mu(x)U_\nu(x+\hat{\mu})U_\mu^\dagger(x+\hat{\nu})U_\nu^\dagger(x)$ is the
plaquette operator.

For the topological charge~\eqref{eq:cont_def_topocharge}, we adopt
the simplest discretization with definite parity, the so-called \emph{clover}
discretization:
\beq\label{eq:clover_charge_def}
Q_{\mathit{clov}} = \frac{1}{2^9\pi^2} \sum_{x,\mu,\nu,\rho,\sigma} \varepsilon_{\mu\nu\rho\sigma} \Tr\left\{C_{\mu\nu}(x)C_{\rho\sigma}(x)\right\},
\eeq
where $C_{\mu\nu}(x)$ is a discretization of the field strength given by the sum
of all the 4 plaquettes centered in the site $x$ and lying on the $\mu$--$\nu$
plane. $Q_{\mathit{clov}}$ is generically non-integer and it is related,
configuration by configuration, to the physical charge $Q$
by~\cite{Campostrini:1988ab}:
\beq\label{eq:lat_topo_charge_renormalization}
Q_{\mathit{clov}} = Z Q + \eta,
\eeq
where $Z$ is a finite renormalization constant that approaches 1 in the
continuum limit, and $\eta$ is a stochastic noise due to ultraviolet (UV)
fluctuations at the scale of the lattice spacing.  Using the variance of
$Q_{\mathit{clov}}$ to estimate the topological susceptibility would require to
take into account both multiplicative and additive renormalizations, which can
be avoided by using one of the several smoothing procedures that have been
proposed in the literature, such as the gradient flow~\cite{Luscher:2009eq,
Luscher:2010iy} or cooling~\cite{Berg:1981nw, Iwasaki:1983bv, Itoh:1984pr,
Teper:1985rb, Ilgenfritz:1985dz, Campostrini:1989dh, Alles:2000sc}, which are
all known to agree with each other when properly matched~\cite{Bonati:2014tqa,
Alexandrou:2015yba}. In this work we use cooling 
due to its simplicity and numerical effectiveness. 

We denote by $Q_{\mathit{clov}}^{\mathit{cool}}$ the topological charge
obtained by measuring the observable Eq.~\eqref{eq:clover_charge_def} on a
configuration to which a certain number of cooling steps have been applied.  To
assign an integer topological charge $Q_L$ to each configuration we follow
Ref.~\cite{DelDebbio:2002xa}, defining
\beq\label{eq:lat_def_topocharge}
Q_L = \round \left\{ \alpha \, Q_{\mathit{clov}}^{\mathit{cool}}\right\},
\eeq
where ``\round'' denotes the rounding to the closest integer and the value of
$\alpha$ is fixed by minimizing
\beq\label{eq:alpha_def}
\braket{ \left( \alpha \, Q_{\mathit{clov}}^{\mathit{cool}} - \round \left\{ \alpha \, Q_{\mathit{clov}}^{\mathit{cool}}\right\}  \right)^2 }\ ,
\eeq
so that the maxima in the distribution of 
$\alpha Q_{\mathit{clov}}^{\mathit{cool}}$ are
located approximately at integer values; such fixing is 
performed at $\theta = 0$ and then adopted also for $\theta \neq 0$. 
The topological susceptibility computed using
$Q_L$ becomes stable (i.e.~independent of the number of cooling steps
$n_{\mathit{cool}}$) after $n_{\mathit{cool}}\sim 10$, 
moreover such threshold 
reveals to be weakly dependent on the lattice spacing, 
thus we chose $n_{\mathit{cool}}=20$ to
define the topological charge in all simulations, verifying also the stability of all continuum extrapolations if a different value of $n_{\mathit{cool}}$ is used.

\subsection{Imaginary-$\theta$ method}
\label{sec:imtheta}

As anticipated in the introduction, the imaginary-$\theta$ method is a
technique that is useful to estimate the topological susceptibility and
especially the coefficients $b_{2n}$ introduced in
Eq.~\eqref{eq:vacuum_energy_theta_dep_parametrization}. In this section we
provide a short summary of this computational method, referring to
Ref.~\cite{Bonati:2015sqt} for more details.
The idea of the method is to introduce an 
imaginary $\theta$ term, in order to avoid a sign
problem, and to extract $\chi$ and $b_{2n}$ from the dependence on $\theta$ of
the cumulants of the topological charge distribution:
the method wins over the standard computation at $\theta = 0$ 
since now the information on the $b_{2n}$ parameters is contained
in all cumulants, including the lowest order ones.
The procedure is most conveniently explained by working 
formally in the continuum: the continuum euclidean action can be written 
in the form
\beq\label{eq:cont_imag_theta_action_def}
S(\theta_I) = S_{\YM} - \theta_I Q,
\eeq
where $\theta_I = i \theta$. The dependence on $\theta_I$ of the cumulants of the
topological charge distribution can be computed from the
derivatives of $E(\theta)$ in Eq.~\eqref{eq:free_energy_def}, properly
continued to the imaginary axis, and can be expressed in terms of $\chi$ and
$b_{2n}$ using Eq.~\eqref{eq:vacuum_energy_theta_dep_parametrization}. As an
example, the explicit expressions of the first few cumulants as a function of
$\theta_I$ read:
\beq\label{eq:cumul_dep_imag_theta}
\begin{split}
\frac{k_1(\theta_I)}{V} &= \chi \left[\theta_I -2b_2 \theta_I^3+3b_4 \theta_I^5+O(\theta_I^6)\right],\\
\frac{k_2(\theta_I)}{V} &= \chi \left[1-6b_2 \theta_I^2+15b_4 \theta_I^4+O(\theta_I^5)\right],\\
\frac{k_3(\theta_I)}{V} &= \chi \left[-12b_2 \theta_I+60b_4 \theta_I^3+O(\theta_I^4)\right],\\
\frac{k_4(\theta_I)}{V} &= \chi \left[-12b_2 +180b_4 \theta_I^2+O(\theta_I^3)\right],
\end{split}
\eeq
where $V$ is the space-time volume and the first fourth cumulants of the topological charge are
\beq\label{eq:first_few_cumuls}
\begin{split}
k_1 &= \braket{Q},\\
k_2 &= \braket{Q^2} - \braket{Q}^2,\\
k_3 &= \braket{Q^3} - 3 \braket{Q^2}\braket{Q} + 2 \braket{Q}^3,\\
k_4 &= \braket{Q^4} - 4 \braket{Q^3}\braket{Q} -3 \braket{Q^2}^2\\
&\quad \, + 12 \braket{Q^2}\braket{Q}^2 - 6 \braket{Q}^4.
\end{split}
\eeq
All these averages are computed by using the weight $e^{-S(\theta_I)}$ in the path-integral.

Let us now describe what changes on the lattice: the lattice action is
\beq\label{eq:lat_imag_theta_action_def}
S_L(\theta_L) = S_W - \theta_L Q_{\mathit{clov}},
\eeq
where the $\theta$-term is discretized by using the non-smoothed clover charge
$Q_{\mathit{clov}}$ defined in the previous subsection, and $\theta_L$ is the
bare imaginary-$\theta$ coupling. The reason for using $Q_{\mathit{clov}}$ is
that with this choice standard heatbath and overrelaxation algorithms can be
used in the update. Relations analogous to
Eq.~\eqref{eq:cumul_dep_imag_theta} can be obtained, where $\theta_I= Z
\theta_L$ and $Z$ is the renormalization constant appearing in
Eq.~\eqref{eq:lat_topo_charge_renormalization}. Measuring the cumulants for several values
of $\theta_L$ we can thus fit the values of $\chi$, $b_{2n}$ and $Z$. We explictly note that 
the cumulants are not affected by the renormalization of $Q_{\mathit{clov}}$
since they are evaluated by using the smoothed and rounded charge $Q_L$ introduced in the previous section (see Ref.~\cite{Berni:2019bch} 
for a more detailed discussion on this point).

\subsection{Parallel tempering of volume defect}
\label{sec:defect}

The lattice action in Eq.~\eqref{eq:lat_imag_theta_action_def}, as the standard
Wilson action, is linear in each of the link variables, hence standard heathbath~\cite{Creutz:1980zw, Kennedy:1985nu} and 
overrelaxation~\cite{Creutz:1987xi}
algorithms can be applied to update the gauge configurations when using the
imaginary-$\theta$ method. However, as anticipated in the introduction, the
local nature of this updating procedure results in a slowing down
of the topological modes, which is exponential
both in the inverse lattice spacing 
and in the number $N$ of colors. This makes very difficult to
perform controlled continuum extrapolations for large values of $N$, since even
simulations with moderately small values of the lattice spacing become
prohibitively expensive as $N$ is increased.

To mitigate this problem, we adopt, in this work, the
parallel tempering algorithm proposed for the $CP^{N-1}$ models in
Ref.~\cite{Hasenbusch:2017unr}, where this algorithm has been shown to perform as well as simulations with open boundaries while bypassing their complications
related to finite-size effects.  Moreover, as shown for $CP^{N-1}$ models in
Ref.~\cite{Berni:2019bch}, parallel tempering can be easily applied in
combination with the imaginary-$\theta$ method discussed in the previous
subsection.

In this algorithm we consider $N_r$ identical systems, differing only for the
boundary conditions on a cuboid defect located on a given spatial slice. In
particular, the boundary conditions imposed on the links that orthogonally
cross the defect are chosen so that the different copies interpolate between
periodic boundary conditions (pbc) and open boundary conditions (obc). Each
system is evolved independently using standard local algorithms, and different
copies are exchanged from time to time with a Metropolis step, so that the
strong reduction of the autocorrelation time achieved in the obc copy is
transferred to the pbc one, on which the measure of the cumulants of the
topological charge $Q_L$ is performed.  Since the injection (or ejection) of
topological charge in the system is mainly triggered by the update of the links
close to the defect, it is convenient \cite{Hasenbusch:2017unr} to alternate
updating sweeps over the whole lattice with hierarchic updates over sub-regions
of the lattice centered around the defect. In particular, we updated more
frequently small hyper-rectangular regions centered around the defect.

In our simulations the location of the defect is the tridimensional region
\beqnn
\begin{aligned}
D=\Big\{&x_1=L-a, \, 0\le x_2 < L_d^{(2)}, \, \\
&0\le x_3 < L_d^{(3)}, \, 0\le x_4 < L_d^{(4)} \Big\},
\end{aligned}
\eeqnn
however, after every hierarchic update, we perform a random translation of the
pbc copy by one lattice spacing, thus effectively moving the location of the
defect. For the sake of the simplicity we use a cubic defect
$L_d^{(2)}=L_d^{(3)}=L_d^{(4)}\equiv L_d$ and it is sufficient to choose $L_d$
equal to a few lattice spacings to obtain satisfactory performances. For a
discussion on how the choice of $L_d$ affects the efficiency of the algorithm,
see Sec.~\ref{sec:algorithm_comparison}.

In order to specify how the different boundary conditions across the defect are
implemented, it is convenient to rescale each link of every replica according
to
\beqnn
U^{(r)}_\mu(x) \to K_\mu^{(r)}(x) U^{(r)}_\mu(x),
\eeqnn
where $U_\mu^{(r)}(x)$ indicates a link of the $r^{\text{th}}$ replica and the
explicit expression of $K^{(r)}_\mu(x)$ is:
\beqnn
K_\mu^{(r)}(x) =
\begin{cases}
c(r), \quad &\mbox{ if} \quad \mu \ne 1 \ \mathrm{and}\ x \in D, \\
1, \quad &\mbox{ otherwise,}
\end{cases}
\eeqnn
so that only the links crossing the volume defect are affected by its presence.
For the pbc replica (corresponding to $r=0$) we have $c(0)=1$, for the obc
replica (corresponding to $r=N_r-1$) we have $c(N_r-1)=0$, for $0<r<N_r-1$ the
value of $c(r)$ interpolates between 0 and 1. With these notations the action
of the $r^{\text{th}}$ copy reads
\begin{align*}
S_L^{(r)}(\theta_L) = & \,\, S_W^{(r)} +S_\theta^{(r)}(\theta_L)\\
= &-\frac{\beta}{N} \sum_{x,\mu>\nu} K^{(r)}_{\mu\nu}(x)\Re \Tr \left\{ \Pi^{(r)}_{\mu\nu}(x)\right\}\\
&- \theta_L Q_{\mathit{clov}}\left[U_\mu^{(r)}(x)\right],
\end{align*}
where $K^{(r)}_{\mu\nu}(x)$ is a short-hand for
\beqnn
K^{(r)}_{\mu\nu}(x) \equiv K^{(r)}_\mu(x) K^{(r)}_\nu(x+\hat{\mu}) K^{(r)}_\mu(x+\hat{\nu}) K^{(r)}_\nu(x).
\eeqnn
Note that we chose to keep the $\theta$ term insensitive to the presence of the
defect, which only affects the Wilson part of the action. Exploratory
simulations performed by modifying also the $\theta$ term provided
evidence that this choice does not significantly affects the performance of the
algorithm, analogously to what was found for $CP^{N-1}$ models in
Ref.~\cite{Berni:2019bch}. This is not surprising since the barriers between
the topological sectors, responsible of the critical slowing down, stem essentially
from the Wilson term.

The swap of replicas was proposed after every step of hierarchic update, for
every couple of adjacent copies $r$ and $r+1$ (differentiating the cases of
even and odd $r$ in order to avoid synchronization problems), and was accepted
with the Metropolis probability
\beq
\begin{aligned}
p =& \min\left\{1, \exp\left\{-\Delta S \right\}\right\}\\
=& \min\left\{1, \exp\left\{- S_W^{\mathit{swap}} + S_W^{\mathit{no\,swap}} \right\}\right\},
\end{aligned}
\eeq
where the $\theta$-term, which is not affected by the defect, does not enter
the acceptance probability. Since boundary conditions of different
replicas differ only on a sub-region of the lattice, to compute $\Delta S$ it
is sufficient to sum the contributions to the action of the plaquettes centered
on sites lying in a hyper-cuboid region centered around the defect and
extending one lattice spacing from it. 

For the parallel tempering to be effective in decorrelating the topological
modes of the pbc copy there must be no bottleneck for a configuration in the
obc copy to be swapped toward the pbc one. In order to guarantee a random walk
without barrier of the configurations between the different replicas it is thus
convenient to choose the constants $c(r)$ entering $K^{(r)}_\mu(x)$ in such a
way that the acceptance ratio is constant for all the proposed swaps:
$p(0, 1)=p(1,2)= \cdots =p(N_r-2,N_r-1)$.

\section{Numerical results}
\label{sec:results}

In order to compare the performances of parallel tempering with the ones of
the standard algorithm, we performed simulations for $N=4$ and $6$ with both
algorithms and for several values of $\beta$ at $\theta_L=0$, measuring the
autocorrelation time of $Q^2_L$. We then performed simulations at non-zero
$\theta_L$ with parallel tempering for each value of $\beta$, to estimate
$\chi$, $b_2$ and $b_4$ using the imaginary-$\theta$ method.

In Tab.~\ref{tab:simulations_summary} we summarize the parameters of the
performed simulations. Lattice sizes have been chosen
to ensure that $L\sqrt{\sigma}\gtrsim 3$, where $\sigma \simeq \left(440 \text{
MeV}\right)^2$ is the string tension, so as to have finite-size effects under
control~\cite{DelDebbio:2001sj,Bonati:2016tvi}.  Statistics reported in
Tab.~\ref{tab:simulations_summary} refer to the parallel tempering simulations
and are reported in number of parallel tempering steps.  A single step of
tempering update consists first of all of
a complete update of each replica, using 5 lattice sweeps of
over-relaxation~\cite{Creutz:1987xi} followed by 1 lattice sweep of
heat-bath~\cite{Creutz:1980zw,Kennedy:1985nu}, both implemented \emph{\`a la}
Cabibbo--Marinari~\cite{Cabibbo:1982zn}, i.e., updating all the $N(N-1)/2$
diagonal $SU(2)$ subgroups of $SU(N)$. After this ``global'' update step we perform 
an iteration on the sub-lattices entering the hierarchical update (see Sec.~\ref{sec:defect}), each iteration consisting of
\begin{itemize}
\item a local update sweep of the sub-lattices for every replica, using the same 
combination of local algorithms adopted for the global update;
\item the parallel tempering swap proposal;
\item a random translation of the pbc copy by one lattice spacing.
\end{itemize}
Since each system is updated using the same procedure and the time required for
hierarchic updates, swap and translations is negligible with respect to the
time of the global update, the total numerical effort of a single parallel
tempering step is $\sim N_r$ times larger than the one required for the local update
in the standard setup.

\begin{table}[!htb]
\centering
\begin{center}
\begin{tabular}{|c|c|c|c|c|c|c|}
\hline
\multicolumn{7}{|c|}{ }\\[-1em]
\multicolumn{7}{|c|}{$N=4$}\\
\hline
&&&&&&\\[-1em]
$\beta$ & $L/a$ & $a\sqrt{\sigma}$ & $L\sqrt{\sigma}$ & $\theta_L^{\mathit{max}}$ & \makecell{Stat.\\$\theta=0$} & \makecell{Stat.\\$\theta\neq0$}\\
\hline
&&&&&&\\[-1em]
11.104 & 16 & 0.1981(5)* & 3.17 & 15 & 255k  & 787k  \\
11.347 & 20 & 0.1590(6) & 3.17 & 15 & 1.39M & 2.44M \\
\hline
\multicolumn{7}{c}{ }\\
\hline
\multicolumn{7}{|c|}{ }\\[-1em]
\multicolumn{7}{|c|}{$N=6$} \\
\hline
&&&&&&\\[-1em]
$\beta$ & $L/a$ & $a\sqrt{\sigma}$ & $L\sqrt{\sigma}$ & $\theta_L^{\mathit{max}}$ & \makecell{Stat.\\$\theta=0$} & \makecell{Stat.\\$\theta\neq0$}\\
\hline
&&&&&&\\[-1em]
24.768 & 12 & 0.2912(11) & 3.57 & 15   & 103k  & 257k  \\
24.845 & 12 & 0.2801(13) & 3.41 & 15   & 113k  & 166k  \\
25.056 & 12 & 0.2499(10) & 3.04 & 15   & 228k  & 280k  \\
25.394 & 14 & 0.2143(8)  & 3.00 & 15   & 513k  & 553k  \\
25.750 & 16 & 0.1878(18) & 3.00 & 17.5 & 1.12M & 1.84M \\
\hline
\end{tabular}
\end{center}
\caption{Summary of simulation parameters. Simulations at non-zero values of
$\theta_L$ were performed in steps of $\Delta \theta_L = 2$ up to $\theta_L=10$
and in steps of $\Delta \theta_L=2.5$ for $\theta_L>10$. The last column refers
to the total statistics accumulated for all imaginary-$\theta$ simulations. The
defect length was, in all cases, $L_d/a=2$. All simulations for $N=4$ were
performed using $N_r=10$, corresponding to a constant swap probability $p$
around $20\%$, while simulations for $N=6$ used $N_r=17$, corresponding to
$p\approx 30\%$.
Lattice spacings are taken from Ref.~\cite{Lucini:2005vg} or
interpolation/extrapolation of data thereof, except for the one marked with *,
which comes from Ref.~\cite{DelDebbio:2001sj}.}
\label{tab:simulations_summary}
\end{table}

\subsection{Parallel tempering: results and comparison}
\label{sec:algorithm_comparison}

\begin{figure}
\centering
\includegraphics[scale=0.5]{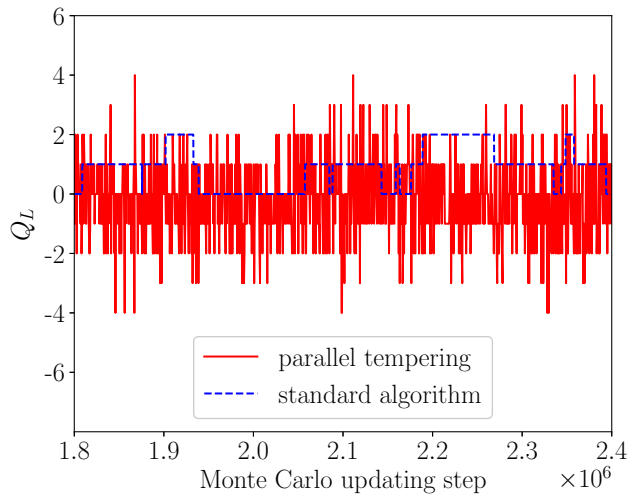}
\caption{Monte Carlo time evolution of the lattice topological charge $Q_L$ for
a run with $N=6$, $\beta=25.75$ and $\theta_L=0$. For the comparison to be
fair, data for the parallel tempering case are plotted as a function of the total number
of global updates performed on all the replicas, i.e. $17$ times the number of
parallel tempering updates.}
\label{fig:Q_MC_evolution}
\end{figure}

Just by inspection of the time histories of the topological charge it is simple to
realize that parallel tempering substantially reduces the topological freezing
allowing us to perform simulations at values of the lattice spacings which
would have been otherwise prohibitive with the standard algorithm. An example
of Monte Carlo time evolution of $Q_L$ obtained with parallel tempering for
$N=6$, $\beta=25.75$ and $\theta_L=0$ is shown in
Fig.~\ref{fig:Q_MC_evolution}, where we compare it with the evolution obtained
with the standard algorithm.

In order to quantitatively characterize the gain achieved with parallel
tempering and optimize its efficiency, it is useful to study the
autocorrelation time of the topological susceptibility.  We use as definition
of the autocorrelation time for the generic observable $\mathcal{O}$ the
expression~\cite{Berg:2004fd} 
\beq\label{eq:def_tau}
\tau(\mathcal{O}) = \frac{1}{2}\left( \frac{\Delta_{\mathcal{O}}^{\mathit{binned}}}{\Delta_{\mathcal{O}}^{\mathit{naive}}} - 1\right)^2,
\eeq
where $\Delta_{\mathcal{O}}^{\mathit{binned}}$ is the error associated to
$\langle \mathcal{O}\rangle$ by using a self-consistent binning analysis, while
$\Delta_{\mathcal{O}}^{\mathit{naive}}$ is the usual standard error of the mean
for independent identically distributed samples. The autocorrelation time of
the topological susceptibility, however, does not take into account the increased
computational effort of the parallel tempering algorithm with respect to the
standard local algorithms. As a figure of merit for the computational effectiveness of parallel 
tempering,  it is thus convenient to introduce the effective autocorrelation time given by
\beq
\tau_{\mathit{pt}}(\mathcal{O})=\tau(\mathcal{O})N_r\ .
\eeq

\begin{figure}[!htb]
\centering
\includegraphics[scale=0.5]{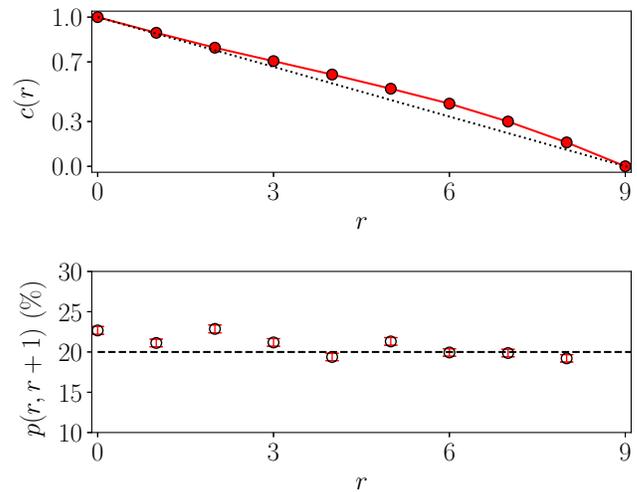}
\caption{Behavior of $c(r)$ as a function of the replica index $r$
compared to a simple linear behavior (figure above), along with the
corresponding acceptances ($\sim20\%$) for the swap between copies $r$ and
$r+1$ (figure below). Data refer to a run with $N=4$, $\beta=11.347$
and $\theta_L=0$.}
\label{fig:acceptances}
\end{figure}

\begin{figure}[!htb]
\centering
\includegraphics[scale=0.5]{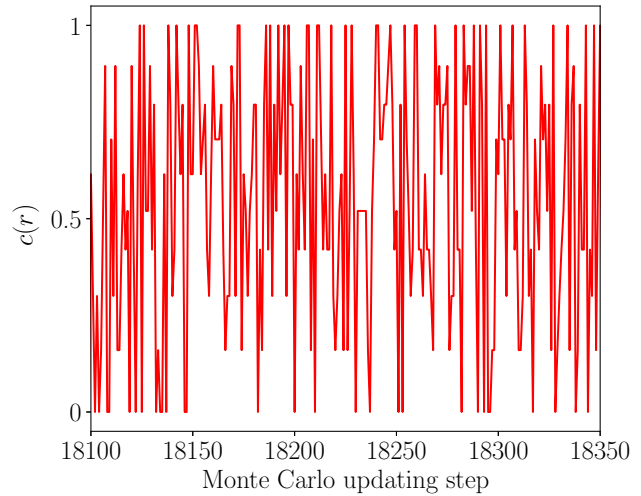}
\caption{Random walk of a configuration among different replicas during a parallel tempering run for $N=4$, $\beta=11.347$ and $\theta_L=0$. Replicas are parametrized by their value of $c(r)$ and the Monte Carlo time is expressed in units of the parallel tempering update step defined in the text.}
\label{fig:bound_cond_MC_evolution}
\end{figure}

As discussed in the previous section, in every parallel tempering simulation we
tuned the parameters $c(r)$ in such a way that the acceptance $p(r,r+1)$ of the
Metropolis swap move between the replicas $r$ and $r+1$ is approximately
independent of $r$. This tuning was performed using test simulations at
$\theta_L=0$ (acceptances do not depend on $\theta_L$), and in
Fig.~\ref{fig:acceptances} we show an example of the behavior of $c(r)$ for a
run with $N=4$ and $\beta=11.347$. Deviations from the linear behavior
appear to be small in the optimal $c(r)$ values, however using these values
$\tau_{\mathit{pt}}(Q_L^2)$ is about half the one obtained by using a
simple linear interpolation. Once $p(r,r+1)$ is almost independent of $r$,
configurations move freely among the different replicas following a random
walk, as shown in Fig.~\ref{fig:bound_cond_MC_evolution}.

The constant value $p$ that is reached by $p(r,r+1)$ after the tuning of $c(r)$
obviously depends on the number $N_r$ of replicas used. We did not perform a
systematic investigation of the dependence on $p$ of the numerical
effectiveness of parallel tempering, however this dependence seems to be quite
mild. Indeed by increasing the number of replicas $N_r$, the constant
acceptance probability $p$ grows and the autocorrelation time $\tau(Q_L^2)$ is
reduced, however also the computational cost increases with $N_r$.  The net
effect is that $\tau_{\mathit{pt}}(Q_L^2)$ is largely insensitive to the
specific value of $p$, at least as far as it is not too close to $0$ or $1$, as we
verified in some test simulations. For example, simulations performed with
$N=4$ and $\beta=11.104$ using $p\simeq 20\%$ (achieved with $N_r=10$) or
$p\simeq 30\%$ (corresponding to $N_r=12$) provided consistent values
of $\tau_{\mathit{pt}}(Q_L^2)$: $72(10)$ and $78(18)$ respectively.
The value of $p$ was kept fixed while approaching
the continuum limit, and in all cases we found that this could be achieved by using the same 
value of $N_r$ for the different lattice spacings (at fixed physical volume).


Most of the simulations reported in this work used $L_d/a=2$ for the size of the defect, 
a value that is sufficient to drastically reduce the freezing problem.
To investigate the dependence of $\tau_{\mathit{pt}}$ on $L_d$ some more simulations
have been performed with $L_d/a=3$ and $L_d/a=1$ for the case $N=6$, always keeping the 
swap acceptance probability $p$ fixed to about $30\%$, which requires to scale
the number of replicas approximately as $\sim\sqrt{L_d^3}$.

\begin{table}[!htb]
\centering
\begin{center}
\begin{tabular}{|c|c|c|c|c|c||c|}
\hline
\multicolumn{7}{|c|}{ }\\[-1em]
\multicolumn{7}{|c|}{$N=4$}\\
\hline
&&&&&&\\[-1em]
$\beta$ & $a\sqrt{\sigma}$ & $L_d/a$ & $p(\%)$ & $N_r$ & $\tau_{\mathit{pt}}\left(Q_L^2\right)$ & $\tau_{\mathit{std}}\left(Q_L^2\right)$ \\
&&&&&&\\[-1em]
\hline
&&&&&&\\[-1em]
\multirow{2}{*}{$11.104$}&
\multirow{2}{*}{$0.1981(5)$} &
\multirow{2}{*}{$2$}
  & 20 & 10 & 72(10) & \multirow{2}{*}{140(10)*} \\
&&& 30 & 12 & 78(18) &                           \\
\hline
&&&&&&\\[-1em]
11.347 & 0.1590(6) & 2 & 20 & 10 & 380(80) & 1000(200)\\
\hline
\multicolumn{7}{c}{ }\\
\hline
\multicolumn{7}{|c|}{ }\\[-1em]
\multicolumn{7}{|c|}{$N_c=6$}\\
\hline
&&&&&&\\[-1em]
$\beta$ & $a\sqrt{\sigma}$ & $L_d/a$ & $p(\%)$ & $N_r$ & $\tau_{\mathit{pt}}\left(Q_L^2\right)$ & $\tau_{\mathit{std}}\left(Q_L^2\right)$ \\
&&&&&&\\[-1em]
\hline
&&&&&&\\[-1em]
24.768 & 0.2912(11) & 2 & 30 & 17 & 16(3) & 110(10)* \\
\hline
&&&&&&\\[-1em]
\multirow{3}{*}{$24.845$}&
\multirow{3}{*}{$0.2801(13)$} &
1 & \multirow{3}{*}{30} & 7  & 120(30)      & \multirow{3}{*}{220(30)*} \\
&& 2 &                       & 17 & 22(5)   &                           \\
&& 3 &                       & 29 & 30(6)   &                           \\
\hline
&&&&&&\\[-1em]
25.056 & 0.2499(10) & 2 & 30  & 17 & 39(8)   & 800(100)*  \\
\hline
&&&&&&\\[-1em]
25.394 & 0.2143(8)  & 2 & 30  & 17 & 110(40) & 5000(1500) \\
\hline
&&&&&&\\[-1em]
\multirow{2}{*}{$25.750$}&
\multirow{2}{*}{$0.1874(8)$}
& 2 & \multirow{2}{*}{30} & 17 & 760(200) & \multirow{2}{*}{$\sim10^5$**} \\
&&3 &                     & 30 & 43(11)   &                           \\
\hline
\end{tabular}
\end{center}
\caption{Results for the autocorrelation time of $Q^2_L$ obtained by using the
standard and the Hasenbusch algorithm.  Quantities denoted with $*$ are taken
from Ref.~\cite{Bonati:2016tvi}, where the procedure used for the local update
was the same of the present work. The result denoted by ** is just a rough
estimate, since it was impossible to obtain a reliable value even after
$\sim2.5$M trajectories.
}
\label{tab:summary_autocorr_times}
\end{table}

\begin{figure}
\centering
\includegraphics[scale=0.5]{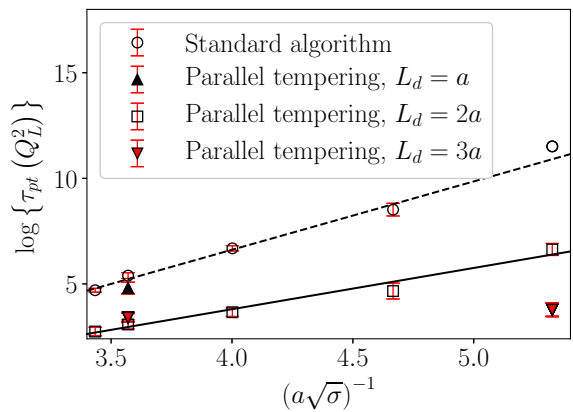}
\caption{Scaling of $\tau_{\mathit{pt}}\left(Q^2_L\right)$ with the inverse
lattice spacing obtained by using the local algorithms or parallel tempering
for $N=6$. The scaling of the autocorrelation time obtained with the standard
algorithm and with parallel tempering at fixed $L_d/a=2$ are both compatible
with an exponential scaling in $1/a$ (dashed and solid line).  Best fits
performed with the fit function $\log\{\tau_{\mathit{pt}}\}= k_0 +
k_1/(a\sqrt{\sigma})$ yield $k_1=3.24(20)$ and $1.95(17)$ for the standard and
the parallel tempering updates respectively.  
}
\label{fig:autocorr_times_N_6}
\end{figure}

A complete list of the obtained autocorrelation times
$\tau_{\mathit{pt}}(Q_L^2)$ is reported in
Tab.~\ref{tab:summary_autocorr_times}, where they are also compared with the
results obtained in Ref.~\cite{Bonati:2016tvi} using just local algorithms. The
scaling of $\tau_{\mathit{pt}}(Q_L^2)$ with $1/(a\sqrt{\sigma})$ for the case
$N=6$ is instead shown in Fig.~\ref{fig:autocorr_times_N_6}.

Simulations performed for $N=6$ at $\beta=24.845$ (corresponding to the points
at $1/(a\sqrt{\sigma})\simeq 3.57$ in Fig.~\ref{fig:autocorr_times_N_6}) using
$L_d/a=1,2,3$ show that $L_d/a=2$ is the optimal choice for this value of the
coupling. As can be seen from Fig.~\ref{fig:autocorr_times_N_6},
autocorrelation times extracted from simulations performed at fixed $L_d/a=2$
are much smaller than the corresponding ones obtained from simulation using
local update algorithms, also for smaller values of the lattice spacing.
However $\tau_{\mathit{pt}}(Q_L^2)$ still seems to scale exponentially with the
inverse lattice spacing. This is due to the fact that, by approaching the
continuum limit at fixed $L_d/a$, the size of the defect in physical units is
reduced, and the mechanism of injection of topological charge through the
defect becomes less and less efficient.

If instead $L_d$ is kept fixed in physical units while approaching the continuum
limit, one generically expects a polynomial critical slowing down in
$1/(a\sqrt{\sigma})$. To investigate this point we performed additional
simulations at $\beta=25.75$ using $L_d/a=3$, in order to have at this lattice
spacing a defect of the same physical size as the one corresponding to
$L_d/a=2$ at $\beta=24.845$ (in both the cases $L_d\sqrt{\sigma} \sim 0.56$).
The outcome of this test is that, despite a $\approx 33\%$ reduction of the
lattice spacing, the effective autocorrelation time $\tau_{pt}(Q_L^2)$ is
compatible in the two cases, as reported in
Tab.~\ref{tab:summary_autocorr_times} and shown in
Fig.~\ref{fig:autocorr_times_N_6}. 

These results still do not permit to make
a clear assessment about the scaling and the optimal tuning
of the parallel tempering algorithm towards the continuum limit,
however, altogether they give a strong indication that 
it works exceedingly well, compared to standard algorithms, 
in reducing topological freezing, and that
the best scaling 
is obtained keeping $L_d$
in the range $0.2-0.3$~fm.
All this is
consistent with what is observed in two-dimensional $CP^{N-1}$
models~\cite{Hasenbusch:2017unr,Berni:2019bch}, where the continuum limit is
performed at fixed $L_d / \xi$, i.e. at fixed physical size of the defect.

\subsection{Analytic continuation and continuum limit}

\begin{figure}
\centering
\includegraphics[scale=0.5]{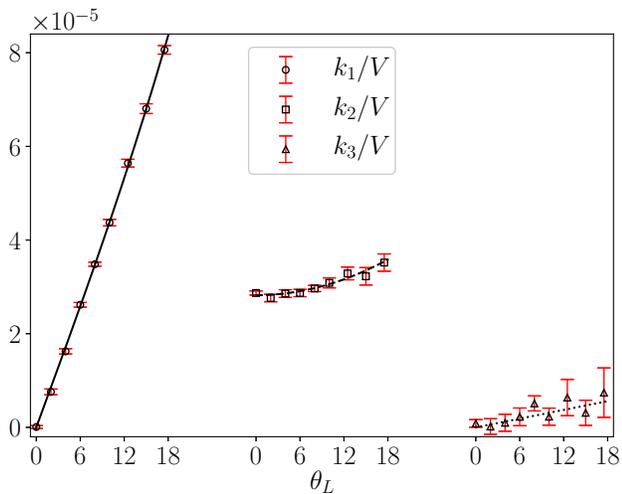}
\caption{Best fit of the first 3 cumulants (solid, dashed and dotted line	respectively) for $N=6$, $\beta=25.750$ and $\theta_L\in \left[0,17.5\right]$, obtained considering up to $O(\theta_L^4)$ terms in the Taylor expansion of the vacuum energy, i.e., neglecting $b_4$ in Eqs.~\eqref{eq:cumul_dep_imag_theta}. The best fit yields $\chi^2/\mathrm{dof}=14.4/24$.}
\label{fig:cumulants_fit_N_6_beta_25.75}
\end{figure}

In Tab.~\ref{tab:summary_results} we summarize the results obtained for the
topological observables $\chi$ and $b_2$ at different values of $N$
and of the lattice spacing, obtained by fitting the $\theta$ dependence of the
cumulants as described in Sec.~\ref{sec:imtheta}. An example of
imaginary-$\theta$ fit of the cumulants is shown in
Fig.~\ref{fig:cumulants_fit_N_6_beta_25.75} for the case $N=6$ and
$\beta=25.75$.

In all the cases we found sufficient to fit the first three
cumulants, as the addition of the fourth one did not change 
the obtained results. Moreover, in all cases we found the $O\left(
\theta_L^6 \right)$ term in the expansion of the vacuum energy to be well compatible with zero since no signal above zero is observed for $b_4$. In particular, we find $\vert b_4(N=4)\vert \cdot 10^5 \lesssim 15$ and $\vert b_4(N=6) \vert \cdot 10^5 \lesssim 30$. For this reason, results for $a^4 \chi$ and $b_2$ reported in Tab.~\ref{tab:summary_results} have been obtained by neglecting $b_4$ in Eqs.~\eqref{eq:cumul_dep_imag_theta}.
Finally we note that correlations between the different cumulants are small and
do not significantly affect the result of the fit, as we explicitly checked by
performing both correlated and uncorrelated fits. 

\begin{table}[!htb]
\centering
\begin{center}
\begin{tabular}{|c|c|c|c|c|}
\hline
\multicolumn{5}{|c|}{ }\\[-1em]
\multicolumn{5}{|c|}{$N=4$}\\
\hline
&&&&\\[-1em]
$\beta$ & $a\sqrt{\sigma}$ & $Z$ & $a^4\chi \cdot 10^5$ & $b_2 \cdot 10^4$\\
\hline
&&&&\\[-1em]
11.104 & 0.1981(5) & 0.14742(96) & 4.183(28) & -137.0(7.5) \\
11.347 & 0.1590(6) & 0.1747(23)  & 1.691(22) & -128(11)    \\
\hline
\multicolumn{5}{c}{ }\\
\hline
\multicolumn{5}{|c|}{ }\\[-1em]
\multicolumn{5}{|c|}{$N=6$} \\
\hline
&&&&\\[-1em]
$\beta$ & $a\sqrt{\sigma}$ & $Z$ & $a^4\chi \cdot 10^5$ & $b_2 \cdot 10^4$ \\
\hline
&&&&\\[-1em]
24.768 & 0.2912(11) & 0.10300(52) & 18.371(93) & -77.2(8.3) \\
24.845 & 0.2801(13) & 0.10945(68) & 15.205(94) & -73.6(9.1) \\	
25.056 & 0.2499(10) & 0.12053(88) & 9.719(69)  & -72.9(8.6) \\
25.394 & 0.2143(8)  & 0.1382(12)  & 5.120(44)  & -65.1(7.6) \\
25.750 & 0.1878(13) & 0.1518(15)  & 2.816(27)  & -58.7(7.4) \\
\hline
\end{tabular}
\end{center}
\caption{Summary of the results obtained using the imaginary-$\theta$ fit for $N=4$ and $6$ by considering up to $O(\theta_L^4)$ terms in the Taylor expansion of the vacuum energy, i.e., neglecting $b_4$ in Eqs.~\eqref{eq:cumul_dep_imag_theta}.}
\label{tab:summary_results}
\end{table}

\begin{figure}
\hspace*{0.16cm}
\includegraphics[scale=0.485]{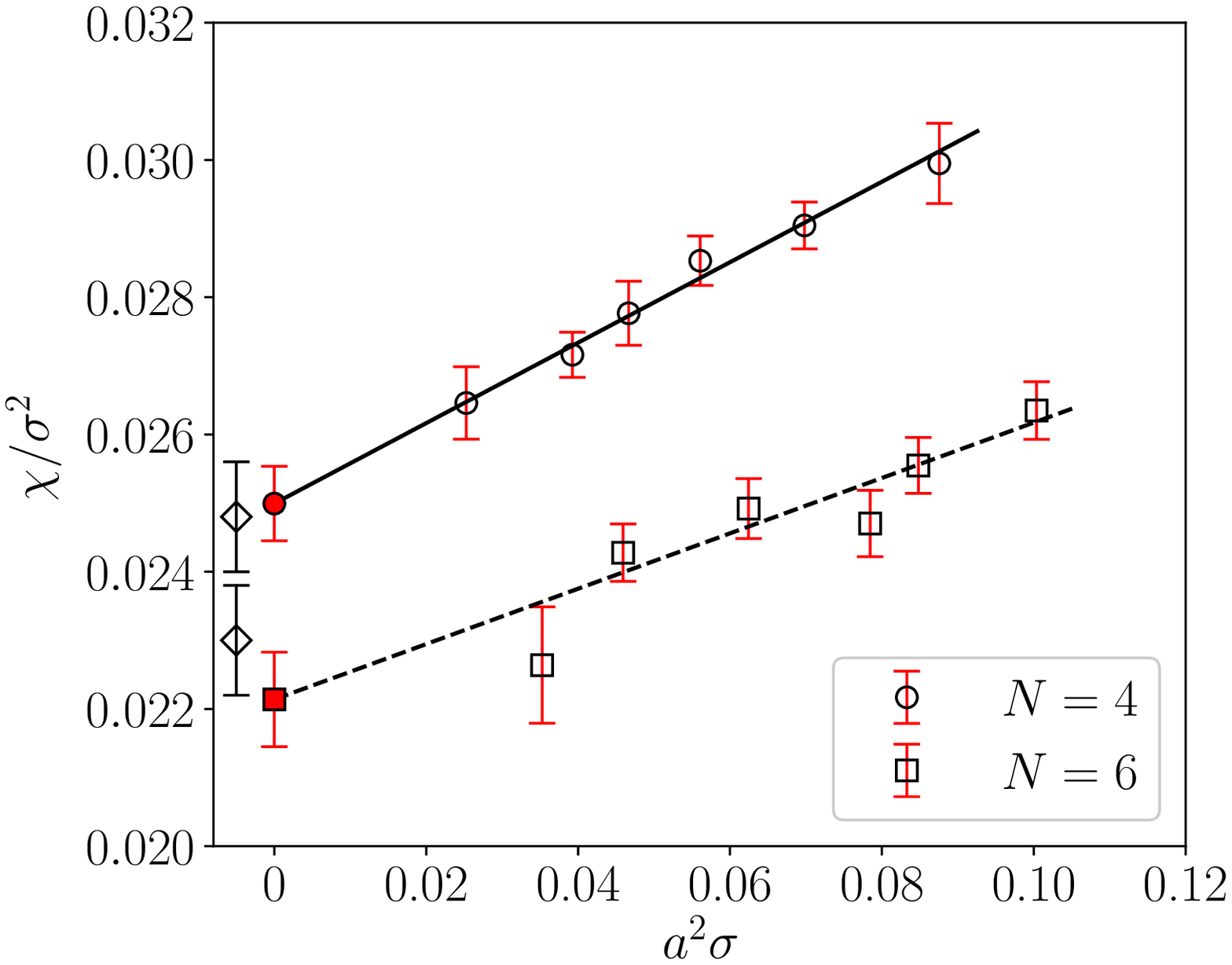}
\hspace*{-0.22cm}
\includegraphics[scale=0.5]{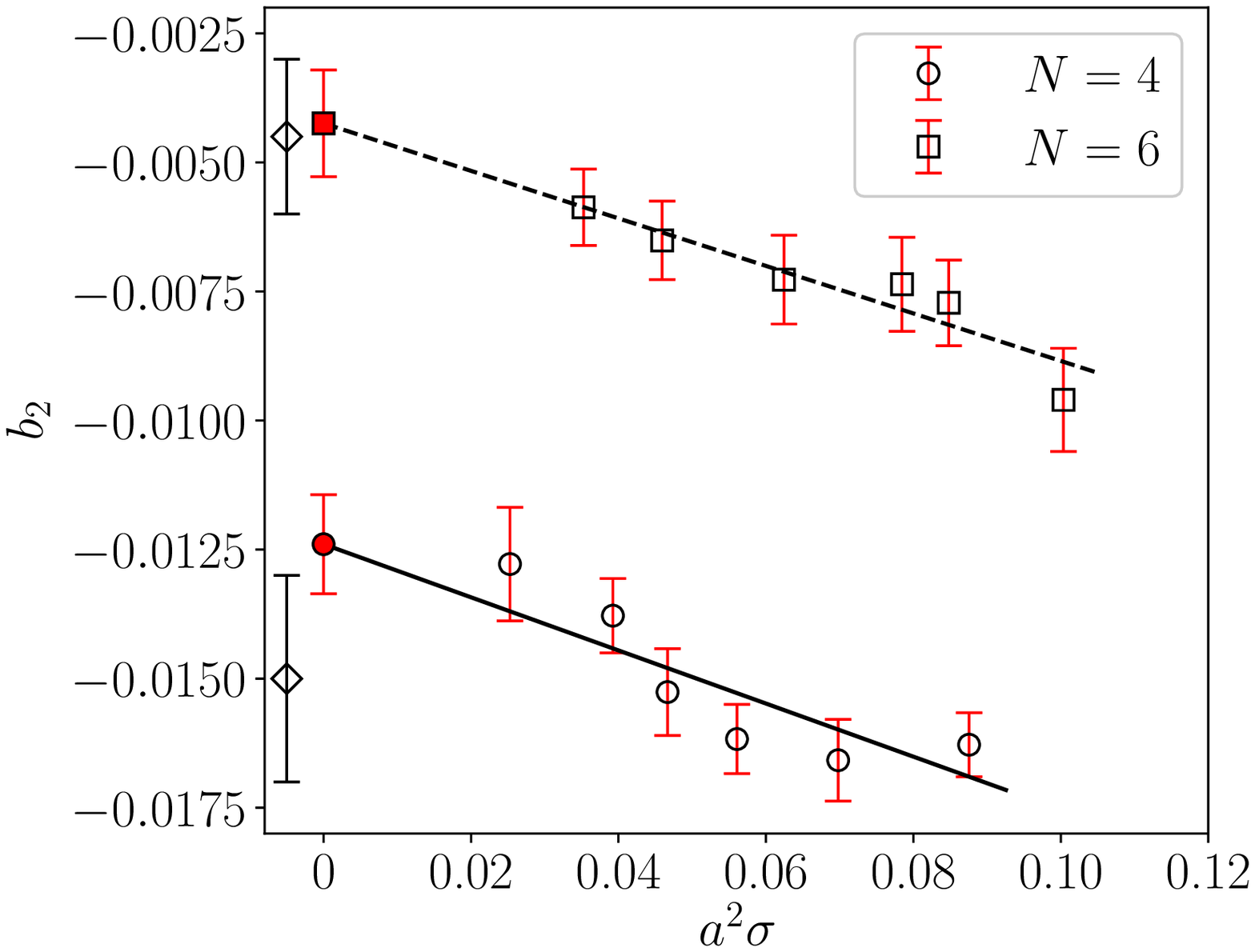}
\caption{Continuum extrapolations of $\chi/\sigma^2$ (above) and $b_2$ (below)
for $N=4$ and $6$ (solid and dashed line respectively) obtained fitting linear corrections
in $a^2 \sigma$ to the continuum limit. The reported best fits yield,
respectively for $N=4$ and $6$, $\chi^2/\rm{dof}=0.8/4$ and $3.7/4$ for
$\chi/\sigma^2$ and $\chi^2/\rm{dof}=4.9/4$ and $1.2/4$ for $b_2$. The diamond
points represent the determinations reported in
Ref.~\cite{Bonati:2016tvi} for $N=4$ and 6.}
\label{fig:cont_limit}
\end{figure}

We used our data for $N=4$ and $N=6$, as well as data obtained for larger
lattice spacings taken from Ref.~\cite{Bonati:2016tvi}, to extrapolate
continuum results for $\chi/\sigma^2$ and $b_2$. For the topological
susceptibility the improvement with respect to previously available
results is only marginal, since the dominant source of error comes from the
string tension $\sigma$ used to set the scale. For $b_2$, instead, we achieved
a substantial improvement of the state of the art, both for $N=4$ and  $N=6$.
In particular for $N=6$ parallel tempering allowed us to reach much finer
lattice spacings than the ones used in previous studies. In this way we could
perform for the first time a controlled continuum extrapolation of $b_2$ in
this case, while in Ref.~\cite{Bonati:2016tvi} only a reasonable
confidence interval was reported. In Tab.~\ref{tab:cont_limit_summary} we
summarize our continuum limits, while in Fig.~\ref{fig:cont_limit} we report
our continuum extrapolations.

\begin{table}[!htb]
\centering
\begin{center}
\begin{tabular}{|c||c|c|}
\hline
&&\\[-1em]
$N$ & $\chi/\sigma^2$ & $b_2$\\
\hline
&&\\[-1em]
3  & 0.0289(13)  & -0.0216(15)  \\
4  & 0.02499(54) & -0.01240(96) \\
6  & 0.02214(69) & -0.0042(10)  \\
\hline
\end{tabular}
\end{center}
\caption{Summary of continuum extrapolations for $N=3,4$ and $6$. Values for $N=3$ are taken from Ref.~\cite{Bonati:2015sqt}.}
\label{tab:cont_limit_summary}
\end{table}

\subsection{Large-$N$ limit}

\begin{figure}[!htb]
\centering
\includegraphics[scale=0.5]{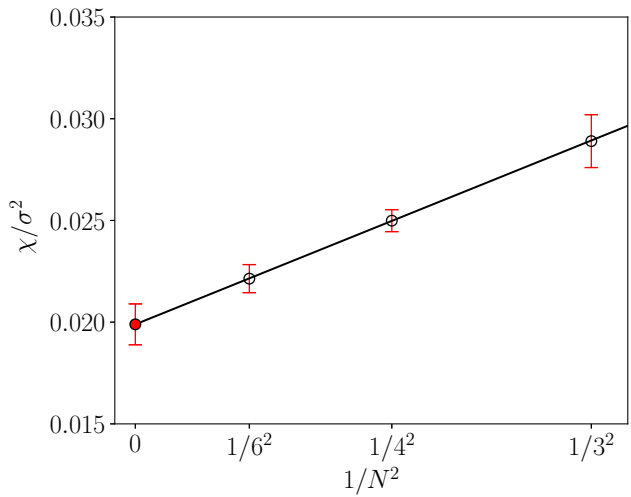}
\caption{Extrapolation of $\chi/\sigma^2$ towards the large-$N$ limit using fit
function $\chi/\sigma^2=\bar{\chi}/\sigma^2+k/N^2$. Best fit yields
$\bar{\chi}/\bar{\sigma}^2=0.0199(10)$ and $k=0.082(17)$.}
\label{fig:large_N_chi}
\end{figure}

In this section we revisit the large-$N$ extrapolation
on the basis of our improved results in particular we report our estimates of $\bar{\chi}$ and $\bar{b}_2$ introduced in Eq.~\eqref{eq:large_N_scaling}.
Let us start from the topological susceptibility: following large-$N$ expectations, we fitted our data for
$N\ge3$ using the functional form:
\beq
\frac{\chi}{\sigma^2} = \frac{\bar{\chi}}{\sigma^2} + \frac{k}{N^2} + O\left(\frac{1}{N^4}\right).
\eeq
Our data are in
agreement with the expected large-$N$ scaling and we
find the result $\bar{\chi}/\sigma^2=0.0199(10)$; the best fit is shown in
Fig.~\ref{fig:large_N_chi} together with numerical results. As already
observed, our result does not improve the previous determination
$\bar{\chi}/\sigma^2=0.0209(11)$ of Ref.~\cite{Bonati:2016tvi} as the main
source of errors comes from the string tension used to set the scale. Using
$\Lambda_{\mathit{large-}N}/\sqrt{\sigma}=0.525(2)$~\cite{GonzalezArroyo:2012bh}
and
$\Lambda_{\mathit{large-}N}=242(10)$~MeV~\cite{Gockeler:2004ad}
to convert to physical units we get
$\bar{\chi}^{1/4}=173(8)$~MeV, in agreement with the prediction
$\bar{\chi}^{1/4}\simeq 180$~MeV obtained from the Witten--Veneziano formula.

\begin{figure}[!htb]
\centering
\includegraphics[scale=0.5]{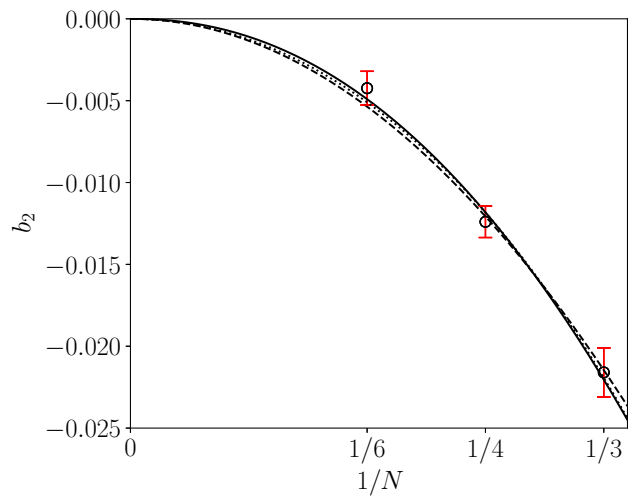}
\caption{Extrapolation of $b_2$ towards the large-$N$ limit. The solid line
represents the best fit obtained using fit function $b_2=\bar{b}_2/N^c$ in the
whole range; the dashed line represents the best fit obtained using
the same fit function but with fixed $c=2$ in the whole range; the dotted line represents the best fit obtained in the whole
range including a further $\bar{b}_2^{\left(1\right)}/N^4$ term in the fit
function. The best fits yield, respectively, $\bar{b}_2=-0.238(79), -0.1931(98)$ and $-0.179(31)$. The free exponent results $c=2.17(26)$, while the next-to-leading correction is $\bar{b}_2^{\left(1\right)}=-0.17(35)$.}
\label{fig:large_N_b2}
\end{figure}

We now pass to the discussion of the large-$N$ behavior of $b_2$. According to
the standard large-$N$ arguments, we expect a behavior of the type:
\beq
b_2 = \frac{\bar{b}_2}{N^2} + \frac{\bar{b}_2^{\left(1\right)}}{N^4} + O\left(\frac{1}{N^6}\right)\ .
\eeq
To test this prediction we perform a best fit of our data with $N\ge3$ using
the power law $b_2(N)= \bar{b}_2/N^{c}$, obtaining a perfect agreement with
expectations, since the exponent results $c=2.17(26)$, which improves the
previous result $c=2.0(4)$ reported in Ref.~\cite{Bonati:2016tvi}. The obtained best fit is shown in Fig.~\ref{fig:large_N_b2}.
By fixing the exponent $c=2$ and fitting our
data with just the leading behavior $b_2=\bar{b}_2/N^2$ in the ranges $N\ge3$
and $N\ge 4$, we obtain the results $\bar{b}_2=-0.1931(98)$ and
$\bar{b}_2=-0.192(14)$ respectively. Since the curve profiles obtained in these two cases are practically indistinguishable, we only show the former in Fig.~\ref{fig:large_N_b2}.
As our final result, we quote the value
$\bar{b}_2=-0.193(10)$, which
improves on the previous determination
$\bar{b}_2=-0.23(3)$ of Ref.~\cite{Bonati:2016tvi}.

\section{Conclusions}
\label{sec:conclusion}

In this work we investigated the $\theta$-dependence of $SU(N)$ Yang--Mills
theories at zero temperature using the parallel tempering algorithm proposed by
Hasenbusch in Ref.~\cite{Hasenbusch:2017unr}. This algorithm was originally
tested in two dimensional $CP^{N-1}$ models, and a first extension of the
original proposal has already been performed in Ref.~\cite{Berni:2019bch}
(still for $CP^{N-1}$ models), by extending the parallel tempering approach to
simulations at imaginary $\theta$ values. In the present work we implemented
the same setup in $SU(N)$ Yang--Mills theory, thus proving the feasibility of
the approach also for computationally more demanding models, and improving the state of the art results for the $\theta$ dependence of these models in the large $N$ limit.

The idea of the method is to simulate many independent identical systems differing only
for the boundary conditions imposed on a cubic defect $L_d\times L_d \times
L_d$, which are chosen to interpolate between open (obc) and periodic boundary
conditions (pbc).  Each replica evolves independently, and swaps among them are
proposed from time to time in order to transfer configurations from the obc to
the pbc replica.  In this way a drastic reduction of the autocorrelation time
of the topological charge is achieved, while avoiding the complication related
to the breaking of translation invariance connected to the adoption of open
boundary conditions, since measures are performed on the pbc replica. 

By using the parallel tempering algorithm we got an impressive reduction of the
autocorrelation times of topological observables, which for the smallest
lattice spacing used was of at least two order of magnitude when taking into
account also the larger computational complexity. A nice feature of the
algorithm is that this gain was obtained without optimally tuning all the
possible parameters entering the update, which proves the robustness of the
approach.  The most relevant parameter to be fixed is clearly the size of the
defect, and we verified that for the cases studied in this paper 
a size in the range $0.2$--$0.3$~fm is sufficiently close to optimal
to obtain a huge reduction of
the critical slowing down. 

The possibility of performing simulations at smaller lattice spacing than in
previous studies
allowed us to achieve a
substantial
improvement in the determination of $\theta$-dependence
beyond the leading $O(\theta^2)$ order. In particular, we 
improved
the accuracy of the determination of the coefficients $b_2$ for both $N=4$ and
$N=6$, in the last case also performing
a controlled continuum limit extrapolation,
which was not 
possible with previously available results.
These data confirm that $b_2$ scales with the
number of colors in a way that is consistent with the leading behavior
predicted by large-$N$ arguments:  data for $N\ge 3$ are  perfectly compatible
with a scaling of the form $b_2 = \bar{b}_2/N^2$, with
$\bar{b}_2=-0.193(10)$,
while a best fit 
according to $b_2=\bar{b}_2/N^c$ returns 
the value $c = 2.17(26)$ for the free exponent.
This shows that the scaling of our data is consistent
with the leading expected behavior, and it is thus reasonable to neglect
sub-leading corrections. We however explicitly note that the accuracy 
is
still
not
high enough to exclude possible unconventional scenarios like the one
put-forward in Ref.~\cite{Vonk:2019kwv}, 
or to better investigate if critical corrections emerge 
at small $N$~\cite{Kitano:2020mfk}. 

A further refinement of present results, including also 
new values of $N$, would thus be welcome in the future: 
the algorithm proposed in Ref.~\cite{Hasenbusch:2017unr},
and extended to $SU(N)$ gauge theories in this study, will 
permit such systematic refinement.

\acknowledgments

Numerical simulations have been performed on the MARCONI machine at CINECA,
based on the agreement between INFN and CINECA (under projects INF19\_npqcd and
INF20\_npqcd).

\end{document}